# Smectic Polymer Vesicles

Lin Jia,[†, ††] Amin Cao,[‡] Daniel Lévy,[†] Bing Xu,[†] Pierre-Antoine Albouy,[§] Xiangjun Xing,[#],

Mark J. Bowick[#] and Min-Hui Li[*†]

[†]Institut Curie, CNRS, Université Pierre et Marie Curie, Laboratoire Physico-Chimie Curie, UMR168, 26 rue d'Ulm, 75248 Paris CEDEX 05, France, [††]Fondation Pierre-Gilles de Gennes pour la Recherche, 29 rue d'Ulm, 75005 Paris CEDEX 05, France, [‡]Laboratory for Polymer Materials, Shanghai Institute of Organic Chemistry, Chinese Academy of Sciences, 354 Fenglin Road, Shanghai 200032, China, [§] Université Paris-Sud, CNRS, Laboratoire de Physique des Solides, UMR8502, 91405 Orsay CEDEX, France, and [#]Physics Department, Syracuse University, Syracuse NY 13244-1130, USA .

[*] Corresponding author:

E-mail: min-hui.li@curie.fr, Tel.: 33 1 56246763, Fax: 33 1 40510636

**Graphical Abstract**

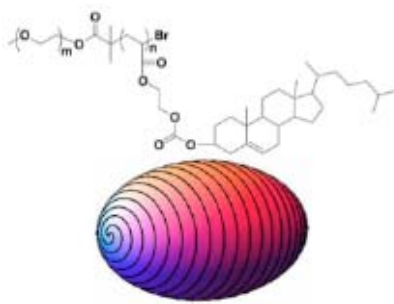

Ellipsoidal smectic polymer vesicles were formed from amphiphilic block copolymers in which the hydrophobic block is a smectic liquid crystal polymer.

**Summary**

Polymer vesicles are stable robust vesicles made from block copolymer amphiphiles. Recent progress in the chemical design of block copolymers opens up the exciting possibility of creating a wide variety of polymer vesicles with varying fine structure, functionality and geometry. Polymer vesicles not only constitute useful systems for drug delivery and micro/nano-reactors but also provide an invaluable arena for exploring the ordering of matter on curved surfaces embedded in three dimensions. By choosing suitable liquid-crystalline polymers for one of the copolymer components one can create vesicles with smectic stripes. Smectic order on shapes of spherical topology inevitably possesses topological defects (disclinations) that are themselves distinguished regions for potential chemical functionalization and nucleators of vesicle budding. Here we report on glassy striped polymer vesicles formed from amphiphilic block copolymers in which the hydrophobic block is a smectic liquid crystal polymer containing cholesteryl-based mesogens. The vesicles exhibit two-dimensional smectic order and are ellipsoidal in shape with defects, or possible additional budding into isotropic vesicles, at the poles.

**Introduction**

Vesicles are membrane-enclosed sacs. Vesicles made of lipids are found frequently in nature and are essential ingredients of biological cells.[1] Lipid vesicles, and their small molecular-weight counterparts, are under intense exploration as potential vessels for drug delivery.[2,3] These vesicles are, however, very fragile because of their essentially liquid character. Polymer vesicles in which the lipids are replaced by amphiphilic block copolymers,[4] are particularly promising. Not only are they tougher and more stable than lipid vesicles,[5] but also their physical, chemical, and biological properties can be tailored by varying the lengths[5] and/or chemical structures[5,7-9] of the polymer blocks, or by conjugation with biomolecules.[10,11] Polymer vesicles are therefore expected to have wider applications in materials science and biotechnology.

There is a close connection between the morphologies of polymer vesicles and the chemical structures of the constituent block copolymers.[4,12] For the case of coil-coil block copolymers, such as poly(acrylic acid)-*b*-polystyrene (PAA-*b*-PS)[6,13] and poly(ethylene oxide)-*b*-polybutadiene (PEO-*b*-PBD),[14] for example, the morphology of polymer vesicles is typically determined by the respective lengths of the polymer blocks, their relative affinity for each other and for the solvents, and physical parameters such as the temperature and the ionic concentration in the case of charged systems. By contrast, for the case of rod-coil block copolymers, the shape anisotropy and additional order in the rod-like block significantly influence the formation and the structures of the resultant supra-macromolecular assemblies. Shape anisotropy and additional order can be introduced by crystalline[15,16] and liquid crystalline (LC) structures[17,18] of the rod-like block, or by secondary structures such as α-helix or β-sheet[19,20] in the case of a peptide. In this report we describe polymer vesicles (Figure1) formed by a smectic block copolymer, PEG-*b*-PAChol, whose chemical structure is given in Figure 2. The mesogenic unit in the LC block PAChol contains a cholesteryl group; the polymer PAChol in its pure form exhibits a thermotropic smectic A phase. The resultant polymer vesicles are found to be ellipsoidal in shape and possess in-plane stripe order, with possible budding into isotropic lobes at the poles-see Figure 1.

The biological world exhibits membranes with a wide variety of shapes and orders. Tubules and buds are two important examples of shapes found in cellular and subcellular processes and their formation has been extensively investigated theoretically.[21-24] It has been shown, for example, that the coupling of the molecular tilt order of the lipids (relative to the normal of the membrane) to the Gaussian curvature favours cylindrical shapes for vesicles.[21,24] Nevertheless, only recently has it become possible to visualize experimentally tilt-ordered domains on the surface of self-assembled lipid tubules.[25] Here we present, for the first time, the direct imaging of smectic order in the membrane of polymer vesicles.

**Results and Discussion**

Amphiphilic LC diblock copolymers were prepared with two different chain lengths and with two fractions of hydrophobic blocks, 72 wt% (Copo1) and 80 wt% (Copo2), by a typical atom transfer radical polymerization as described previously[18] (see Experimental section for details). The hydrophilic block PEG was the same ($M_n$ = 2000, Degree of polymerization m = 45) in both systems. The degrees of polymerization for LC blocks are n = 10 for Copo1 and n = 16 for Copo2 (Figure 2). To determine the LC properties of the hydrophobic block separately, we have also synthesized the corresponding LC homopolymer PAChol ($M_n$ = 5230, n = 10, $M_w/M_n$=1.11, $M_w$ and $M_n$ being respectively weight and number average molecular weights). It possesses a smectic A phase, as indicated by the fan textures (Figure 3a) and also confirmed by the small angle X-ray scattering (SAXS) pattern with two orders of Bragg reflection located at $q$ and $2q$ ($q = 4\pi\sin\alpha/\lambda = 2\pi/P$ is the wave vector, where $2\alpha$ is the scattering angle) (Figure 3b). From these data it was deduced that the smectic layer spacing is P = 4.29 nm, which is between $l$ and $2l$, $l$ = 2.65 nm being the extended mesogen length estimated by Dreiding models. Therefore the mesophase is an interdigitated smectic A phase ($SmA_d$),[26] as shown in Figure 3c. The complete phase sequence of the homopolymer PAChol is: g-68°C-$SmA_d$ 156°C-I, as determined by differential scanning calorimetry (DSC) at 10°C.min$^{-1}$; see Supporting Figure S4 for curves.

Self-assemblies of Copo1 and Copo2 in dilute water solution were performed with the aid of the co-solvent dioxane to fluidize the hydrophobic blocks ($T_g$ = 68°C). The block copolymers were first dissolved in dioxane at a concentration of 1 wt% at 25°C. Deionized water was then added very slowly to the solution, with slight shaking, up to a water concentration of about 50 wt%. The turbid suspensions were then dialyzed against water to remove all dioxane. The final water suspensions deposited on a microscope grid were then fast frozen in liquid ethane, and examined by cryo-transmission electron microscopy (cryo-TEM).

Polymer vesicles formed for both block copolymers were studied. Figure 1 shows clearly that all vesicles are unilamellar and no multilamellar structures are visible. For Copo1 (short copolymer), most of the vesicles are ellipsoidal with long axis less than 300 nm (nano- vesicles), as illustrated in

Figure 1a. A few of the vesicles have long axes up to 1 μm (see Supporting Figure S6) with a round bud at the end (Figure 1b). In contrast, most vesicles formed by Copo2 (long copolymer) are giant vesicles with ellipsoidal shapes (Figure 1c, d). Their long and short axes are typically $D_1 \sim$ 2-5 μm and $D_2 \sim$ 0.3-2 μm (see Supporting Figure S7). A lot of them have spherical buds emanating from the poles (Figure 1d), and others have smooth tips (Figure 1c). Only a few of the vesicles have linear size less than 300 nm. See also Figure S6 and S7 in Supporting Information for cryo-TEM images of Copo1 and Copo2. Note that in contrast the nematic polymer vesicles studied previously, where the hydrophobic block is a side-on nematic polymer, were always spherical.[17,27]

A key question is whether the smectic structure observed in the bulk phase of hydrophobic building block also exists in polymer vesicles. If yes, how is it organised relative to the vesicle membrane? As shown in Figure 1a-d, the obtained images actually show uniform stripes perpendicular to the major axis of the ellipsoidal vesicles for both copolymers. The Fourier transform patterns shown in the insets reveal a periodic spacing of $P = 4.3 \pm 0.1$ nm for the stripes, matching very closely the lamellar spacing of the $SmA_d$ phase of the corresponding homopolymer PAChol, see Figure 3b and c. Thus smectic structure indeed persists in the membrane of polymer vesicles. See Figure 4a for a schematic of the molecular organisation throughout the membrane. Also note that bending of the membrane parallel to the smectic layer doesn't change the layer spacing, while bending perpendicular to the layers makes the layer spacing on both sides of the membrane unequal, and therefore costs extra elastic free energy. This explains why the stripes are always perpendicular to the major axis of vesicles. In conclusion, an in-plane thermotropic smectic LC structure is clearly observed for the first time in the lyotropic membrane of vesicles.

We next discuss the vesicle size and the membrane thickness. The membrane thickness is not uniform everywhere. For Copo1 (n = 10), the thickness falls in the range $e = 8 - 13$ nm in smectic regions, and falls in the range $e = 4 - 7$ nm in the buds where no stripes are visible. For Copo2 (n = 16), they are $e = 10 - 13$ nm and $e = 5 - 7$ nm respectively. Note that cryo-TEM is only sensitive to the hydrophobic part of the membrane. Therefore the hydrophobic thickness of the membrane is very similar for both copolymers in spite of the fact that the LC block chain length of Copo2 (n = 16) is on average 60% longer than that of Copo1 (n = 10). The conformation of hydrophobic chains must therefore be very different in each case.

Vesicle-forming amphiphilic block copolymers in water energetically prefer a planar disk-like bilayer over micelles. Closed vesicles then form when the line energy of a circular rim of bilayer, $E_{disk} = 2\pi R_D \gamma$, where $\gamma$ is the line tension of the rim and $R_D$ the disk radius, exceeds the bending energy needed to form a vesicle, $E_{bend} = 8\pi\kappa$ with $\kappa$ the bending modulus. The resultant minimal vesicle size is then $R_v = 2\kappa/\gamma$.[12] In the present case, vesicle formation is assisted by the dioxane fluidizer. As water is added to the dioxane solution of the copolymer, the solvent mixture becomes an increasingly poorer solvent for the LC blocks. As a result planar bilayer sheets form and subsequently close into vesicles. Generally one expects that the line tension $\gamma$ scales linearly with the membrane thickness e, while the bending modulus $\kappa$ scales as the thickness cubed; hence $R_{V,Copo1}/R_{V,Copo2}$ should scale as $(e_{Copo1}/e_{Copo2})^2$. If we naively take $e_{Copo1}$ and $e_{Copo2}$ to be the thickness of the membranes **after** the dioxane is removed, which are roughly equal (values from cryo-TEM), we would deduce that the vesicle sizes for two systems should be approximately the same, in strong contrast with the observation that the vesicle size for Copo2 is factor of ten larger than that of Copo1. The membrane thickness before the removal of dioxane is likely very different, therefore, from the ultimate thickness measured. This issue remains a mystery, as we are not able to image the vesicle by cryo-TEM in the presence of dioxane.

Finally we address the structure of stripes near the poles of ellipsoids, where direct imaging is difficult in our experimental setup. Striped, or smectic, order on a surface with spherical topology[28] must exhibit orientational defects of total charge +2,[22,28,29] as required by the Gauss-Bonnet-Poincaré theorem. For the polymer vesicles discussed here, since all the smectic layers are roughly perpendicular to the major axis and each polar region should carry disclination charge +1. Due to the limitations of the cryo-TEM method, we are not able to orient the vesicles end-on to image these polar defects. It is nevertheless straightforward to see that all stripes should generically form helices around the major axis; the special case where all smectic layers form closed circles is possible but unlikely. Consequently, the +1 disclinations around each pole are more appropriately characterized as tightly bound pairs of +1/2 disclinations, as illustrated in Figure 4b.

Most of the giant vesicles possess buds at the poles; see, for example, Figure 1b and d. No striped order is observed in these regions, suggesting that these buds are in the isotropic phase. As can also be seen in Figure 1b and d, the buds are typically connected to the main body of giant smectic vesicles by a narrow neck, where presumably an isotropic-smectic domain wall is located. The budding phenomenon is most likely, therefore, a result of smectic-isotropic phase separation. This phase separation must happen during the removal of dioxane, since in the absence of the fluidizer at room temperature the system is in a glassy state where phase separation is not possible; see the phase sequence of the corresponding homopolymer discussed above, and also Supporting Figure S4.

The formation of necks likely results from the non-equilibrium isotropic-smectic phase separation discussed above. This process is, however, sufficiently slow, that the final state of budded vesicles is a local minimum of the elastic free energy, subject to appropriate kinetic constrains. Consequently we should be able to estimate of the neck size as a function of the elastic moduli. Let us consider the simplified geometry as shown in Figure 4c: to the right of the neck is a round isotropic bud (whose radius does not matter in our calculation). The part of the main smectic vesicle immediately to the left of the neck has the shape of a cone with an opening angle $\theta$. In between the neck has the shape of a hyperboloid characterized by two radii of curvature, $a$ and $b$ respectively. For simplicity, we shall assume that the layer spacing in the smectic side is everywhere constant, and ignore the possibility of freely terminated stripes (dislocations). The neck shape should be determined by the competition between the domain wall energy and the bending energies of the membrane and of the stripes: the former prefers a narrow neck and the latter a wide neck. A simple calculation[30] then shows that the total free energy relevant to this neck geometry is given by

$$F = \pi K \sin\theta \log \frac{L \sin\theta}{a} + 2\pi \gamma_{IS} a + \pi \kappa_\perp \frac{a}{b} + \pi \kappa_\parallel \frac{b}{a} \qquad (1)$$

where $K$ is the bending modulus of stripes in the tangent plane of the membrane, while $\gamma_{IS}$ is the line tension of an isotropic-smectic phase boundary; $\kappa_\parallel$ and $\kappa_\perp$ are the moduli for the bending of the membrane in the direction parallel and perpendicular to the stripes. Minimizing the total free energy over $b$ we find that $b$ is proportional to $a$. Substituting back into Eq. (1) we find that the last two terms reduce to a constant which can be dropped. Minimizing the remaining two terms over the radius of curvature $a$, we find $a = K \sin\theta /(2 \gamma_{IS})$. Let us emphasize that both the parameters $K$ and $\gamma_{IS}$ correspond to the stage where isotropic-smectic phase separation occurs. To calculate the angle $\theta$, we need a more complete understanding of the dynamic process of phase separation.

**Conclusion**

We have found that block copolymers, composed of a smectic hydrophobic block (PAChol) and a hydrophilic block (PEG), self-assemble to ellipsoidal smectic vesicles. The nano- smectic vesicles possess topological defects at the ellipsoidal poles, a circumstance that would allow creation of novel divalent colloids with ligands or other functional groups anchored at the defects cores.[29,31] The giant smectic vesicles possess possible budding into isotropic vesicles in the pole regions. These smectic polymer vesicles offer novel examples of the interplay between orientational/positional order and the curved geometry of a two-dimensional membrane.

**Experimental**

**Synthesis**

The LC monomer, cholesteryl acryloyoxy ethyl carbonate (AChol), was synthesized from cholesteryl chloroformate and 2-hydroxyethyl acrylate as decribed previously.[18] Methoxy poly(ethylene glycol) ($M_n$=2000) was first converted to an atom transfer radical polymerization (ATRP) macro-initiator (PEG$_{45}$-Br) by reaction with 2-bromoisobutyryl bromide.[18] The block copolymers PEG$_{45}$-b-PAChol$_{10}$ (Copo1) and PEG$_{45}$-b-PAChol$_{16}$ (Copo2), were synthesized by ATRP as described previously[19] from PEG$_{45}$-Br and AChol, using Cu$^I$Br as catalyst, *N,N,N',N'',N''*-pentamethyldiethylenetriamine (PMDETA) as ligand and xylene as solvent. The LC homopolymer PAChol was also synthesized by ATRP, using ethyl 2-bromo-2-methylpropionate as initiator, Cu$^I$Br as catalyst, 4,4'-di(*n*-nonyl)-2,2'-bipyridine as ligand and toluene as solvent. See Supplementary Information for ATRP protocols for the block copolymer and homopolymer syntheses.

**Characterization**

The chemical structures of all products and molecular weights of the polymers were analyzed by $^1$H-NMR using a Varian VXR 300 FT-NMR spectrometer. LC monomer AChol $^1$H NMR (CDCl$_3$, δ ppm): 0.67-2.42 (m, 43H, -C*H*$_3$, -C*H*(CH$_3$)-, -C*H*-, -C*H*$_2$-), 4.39 (s, 4H, -CO-O-C*H*$_2$-C*H*$_2$-O-CO-), 4.48-4.50 (m, 1H, -C*H*O-), 5.39-5.40 (m, 1H, -C(CH$_2$-)=C*H*-), 5.85-5.90 (m, 1H, -CH=HC*H*), 6.10-6.20 (m, 1H, -C*H*=CH$_2$), 6.42-6.48 (m, 1H, -CH=HC*H*). LC homopolymer PAChol $^1$H NMR (CDCl$_3$, δ ppm): 0.64-2.54 (m, -C*H*$_3$, -C*H*(CH$_3$)-, -C*H*-, -C*H*$_2$-), 4.02-4.13 (q, 2H, -CO-OC*H*$_2$CH$_3$), 4.19-4.42 (s, br, 4*n*H, -CO-O-C*H*$_2$-C*H*$_2$-O-CO-), 4.41-4.62 (s, br, *n*H, -C*H*O-), 5.34-5.46 (s, br, *n*H, C=C*H*-) (*n* is the degree of polymerization of the homopolymer PAChol). PEG$_{45}$-Br $^1$H NMR (CDCl$_3$, δ ppm): 1.95 (s, 6H, -C (C*H*$_3$)$_2$-Br), 3.37 (s, 3H, C*H*$_3$-O-), 3.65 (s, 181H, -O-C*H*$_2$-C*H*$_2$-O-, -O-C*H*$_2$-CH$_2$-OCO-), 4.32-4.35 (m, 2H, -O-CH$_2$-C*H*$_2$-OCO-). Block copolymer PEG-*b*-PAChol $^1$H NMR (CDCl$_3$, δ ppm): 0.67-2.42 (m, -C*H*$_3$, -C*H*(CH$_3$)-, -C*H*-, -C*H*$_2$-), 3.37 (s, 3H, C*H*$_3$-O-), 3.65 (s, 181H, -O-C*H*$_2$-C*H*$_2$-O-, -O-C*H*$_2$-CH$_2$-OCO-), 4.30 (m, (4*n*+2)H, -CO-O-C*H*$_2$-C*H*$_2$-O-CO-, -O-CH$_2$-C*H*$_2$-OCO- ), 4.40-4.50 (m, *n*H, -C*H*O-), 5.37-5.40 (m, *n*H, C=C*H*-). (*n* is the degree of polymerization of the PAChol block). The mean value of *n* for the homopolymer is $n_{\text{PAChol}} = 2 \times (I_{5.34-5.46}/I_{4.02-4.13})$, where $I_{5.34-5.46}$ is the signal integration at 5.34-5.46 ppm, idem for others. The value of *n* for the LC block is $n_{\text{PAChol block}} = 3 \times (I_{5.37-5.40}/I_{3.37})$ and the molecular weight of the diblock copolymer determined by NMR is $M_n = M_{n,\text{PEG}} + n_{\text{PAChol block}} \times 528$, where 528 corresponds to the molecular weight of the LC monomer AChol. See Supplementary Figs. S1, S2 for NMR spectra.

Molecular weight distributions ($M_w/M_n$) of the diblock copolymers were evaluated by size exclusion chromatography (SEC) calibrated with PEG standards. $M_w/M_n$ of the LC homopolymer was evaluated by SEC calibrated with polystyrene standard. We used a Perkin Elmer 200 Series with a chromatography interface NCI 900 and a differential refractometer for SEC. Double columns of PL gel 5 m mixed-D type (300×7.5 mm, *Polymer Laboratory, UK*) were applied in series with chloroform as eluent at a flow rate of 1.0 mL.min$^{-1}$ at 40ºC. Chromatograms of Copo1, Copo2 and their corresponding macroinitiator $PEG_{45}$-Br are given in Supplementary Fig. S3. All molecular weights and molecular weight distributions are summarized in Supplementary Table S1.

The mesomorphic properties of the LC homopolymer were studied by thermal polarizing optical microscopy (POM) using a Leitz Ortholux microscope equipped with a Mettler FP82 hot stage, and differential scanning calorimetry (DSC) using a Perkin-Elmer DSC7. The mesophase structure was studied on fiber samples drawn from molten polymer by X-ray scattering using $CuK_\alpha$ radiation ($\lambda$ = 0.154 nm) from a 1.5 kW rotating anode generator. The diffraction patterns were recorded on photosensitive imaging plates.

**Vesicle preparation**

The preparation of the polymer vesicles and the turbidity measurements were performed at room temperature according to published procedures.[13,27] The diblock copolymers were first dissolved in dioxane, which is a good solvent for both polymer blocks, at a concentration of 1.0 wt%. Deionized water was then added very slowly to the solution under slight shaking. Typically, 2-3 µL of water was added each time to 1 mL of polymer solution, followed by 10 or more minutes of equilibration until the optical density was stable. The optical density (turbidity) was measured at a wavelength of 650 nm using a quartz cell (path length: 2 cm) with a Unicam UV/Vis spectrophotometer. The cycle of water addition, equilibration and turbidity measurement was continued until reaching 50% water concentration. The turbidity diagram is shown in Supplementary Fig. S5. The solution was then dialyzed against water for 3 days to remove dioxane using a Spectra/Por regenerated cellulose membrane with a molecular weight cut-off of 3500.

**Cryo-Electron Microscopy (Cryo-TEM)**

The polymer vesicle suspension was deposited onto a holey grid (Ted Pella Inc., USA) and flash frozen in liquid ethane. The frozen sample was observed with a Philips CM120 electron microscope operating at 120 kV. Images were recorded in low dose mode with a Gatan SSC 1024 x 1024 pixels CCD camera. Calibration was performed with a 2D crystal of purple membrane leading to 0.385 nm/pixel at 45 000 magnification.

**Acknowledgment**. MHL thanks Jacques Prost for helpful advice and Patrick Keller for discussion. MJB thanks David R. Nelson for discussion. We thank Aurélie Di Cicco for obtaining the cryo-TEM images. We acknowledge CNRS and CAS for their support to our Franco-Chinese cooperation. This work also received help from the Agence Nationale de la Recherche (ANR-08-BLAN-0209-01).

**Electronic Supplemental Information Available**: Synthetical protocols for the block copolymer and homopolymer. Supplemental Table S1 for molecular weights and molecular weight distributions of polymers. Supplementary Figure S1-S5: NMR spectra and SEC curves of polymers, turbidity curves of vesicle preparation and cryo-TEM images of polymer vesicles. See http://www.rsc.org/suppdata/xx/b0/b000000x/.

**References**

1. R. Lipowsky and E. Sackmann, Eds., *Structure and Dynamics of Membranes—from Cells to Vesicles*, Elsevier Science, Amsterdam, 1995.

2. D. D. Lasic and D. Papahadjopoulos, Eds., *Medical Applications of Liposomes*, Elsevier Science, Amsterdam, 1998.

3. I. F. Uchegbu, Ed., *Synthetic Surfactant Vesicles: Niosomes and Other Non-Phospholipid Vesicular Systems*, vol . 11 of *Drug Targeting and Delivery*, Harwood Academic, Amsterdam, 2000.

4. D. E. Discher and A. Eisenberg, *Science*, 2002, **297**, 967-973.

5. B. M. Discher, Y. Y. Won, D. S. Ege, J. C.-M. Lee, F. S. Bates, D. E. Discher, D. A. Hammer, *Science*, 1999, **284**, 1143-1146 .

6. T. Azzam and A. Eisenberg, *Angew. Chem. Int. Ed.*, 2006, **45**, 7443-7447.

7. J. C. M. Van Hest, D. A. P. Delnoye, M. W. P. L. Baars, M. H. P. Genderen and E. W. Meijer, *Science*, 1995, **268**, 1592-1595.

8. L. Zhang, K. Yu and A. Einsenberg, *Science*, 1996, **272**, 1777-1779.

9. J. J. L. M. Cornelissen, M. Fischer, N. A. J. M. Sommerdijk and R. J. M. Nolte, *Science*, 1998, **280**, 1427-1430.

10. W. Meier, C. Nardin, and M. Winterhalter, *Angew. Chem. Int. Ed.*, 2000, **39**, 4599-4602.

11. A. Graff, M. Sauer, P. V. Gelder, and W. Meier, *Proc. Natl. Acad. Sci. U.S.A.*, 2002, **99**, 5064-5068.

12. M. Antonietti and S. Förster, *Adv. Mater.*, 2003, **15**, 1323-1333.

13. L. Zhang, and A. Eisenberg, *Science*, 1995, **268**, 1728-1731.

14. S. Jain, and F. S. Bates, *Science*, 2003*,* **300**, 460.

15. S. A. Jenekhe, and X. L. Chen, *Science*, 1998, **279**, 1903-1907; S. A. Jenekhe and X. L. Chen, *Science*, 1999, **283**, 372-375.

16. X. Wang, G. Guerin, H. Wang, Y. Wang, I. Manners and M. A. Winnik, *Science*, 2007, **317**, 644-647.

17. J. Yang, R. Piñol, F. Gubellini, P.-A. Albouy, D. Lévy, P. Keller and M.-H. Li, *Langmuir*, 2006, **22**, 7907-7911.

18. R. Piñol, L. Jia, F. Gubellini, D. Lévy, P.-A. Albouy, P. Keller, A. Cao and M.-H. Li, *Macromolecules*, 2007, **40**, 5625-5627.

19. E. G. Bellomo, M. D. Wyrsta, L. Pakstis, D. J. Pochan and T. J. Deming, *Nature Mater.*, 2004, **3**, 244-248.

20. J. Rodriguez-Hernandez and S. Lecommandoux, *J. Am. Chem. Soc.*, 2005, **127**, 2026-2027.

21. F. C. MacKintosh and T. C. Lubensky, *Phys. Rev. Lett.*, 1991, **67**, 1169-1172.

22. T. C. Lubensky and J. Prost, *J. Phys. II*, 1992, **2**, 371.

23. R. M. L. Evans, *J. Phys. II*, 1995, **5**, 507.

24. H. Jiang, G. Huber, R. A. Pelcovits, T. R. Powers, *Phys. Rev. E*, 2007, **76**, 031908-1-9.

25. Y. Zhao, N. Mahajan, R. Lu, J. Fang, *Proc. Natl. Acad. Sci. U.S.A.*, 2005, **102**, 7438-7442.

26. H. Fischer, S. Poser, M. Arnold and W. Frank, *Macromolecules*, 1994, **27**, 7133-7138.

27. J. Yang, D. Lévy, W. Deng, P. Keller and M.-H. Li, *Chem. Commun.*, 2005, 4345 – 4347.

28. X. Xing, *Phys. Rev. Lett.*, 2008, **101**, 147801; X. Xing preprint *arXiv: 0806.2409*, to appear in the *Journal of Statistical Physics*.

29. D. R. Nelson, Toward a Tetravalent Chemistry of Colloids. *Nano Letters*, 2002, **2**, 1125–1129.

30. This will be discussed in detail in a future theoretical paper on smectic vesicles.

31. G. A. DeVries, M. Brunnbauer, Y. Hu, A. M. Jackson, B. Long, B. T. Neltner, O. Uzun, B. H. Wunsch and F. Stellacci, Divalent metal nanoparticles. *Science*, 2007, **315**, 358-361.

**Figure captions**

**Figure 1** Cryo-transmission electron micrographs of smectic polymer vesicles. (**a**) and (**b**), Copo1. (**c**) and (**d**), Copo2. The scale bars are 100 nm. Insets are Fourier transforms of representative areas of the vesicles. The periodicity of all smectic areas is identical and corresponds to P = 4.3 ± 0.1 nm. The buds of the vesicles in (**b**) and (**d**) are not liquid crystalline as seen on the Fourier transform.

**Figure 2** Amphiphilic LC block copolymers containing a cholesteryl-based mesogen: Copo1 and Copo2. The hydrophilic/hydrophobic weight ratio is 28/72 for Copo1 and 20/80 for Copo2. The schematic on the right shows the hydrophilic PEG in blue connected to the hydrophobic side-chain LC polymer which itself consists of a black backbone and red LC mesogens.

**Figure 3** (**a**) POM image of LC homopolymer PAChol textures at 97°C in the smectic phase. The same texture is preserved up on further cooling to room temperature, where the system becomes a vitreous smectic phase. (**b**) SAXS pattern of PAChol on melt drawing fiber. Two orders of Bragg reflection located at $q$ and $2q$ were observed and a lamellar period of P = 4.29 nm was deduced. $q = 4\pi \sin\alpha/\lambda = 2\pi/P$, is the wave vector. (**c**) Structural model of the $SmA_d$ phase in PAChol fiber. The fiber is along the vertical direction in (**b, c**). The fiber drawing process aligns the smectic layer normal perpendicular to the fiber.

**Figure 4** (**a**) Schematic representation of smectic ellipsoidal polymer vesicles. See Figure 2 for the symbols of cholesteryl mesogens and polymer chains. e is the membrane thickness and P is the smectic period. (**b**) Expected defects structure at a pole of ellipsoid. (**c**) Simplified geometry of a neck.

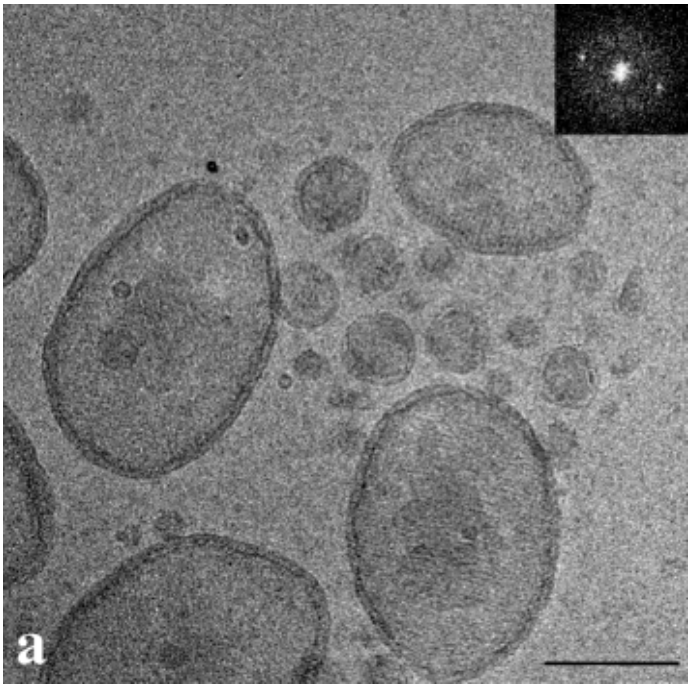 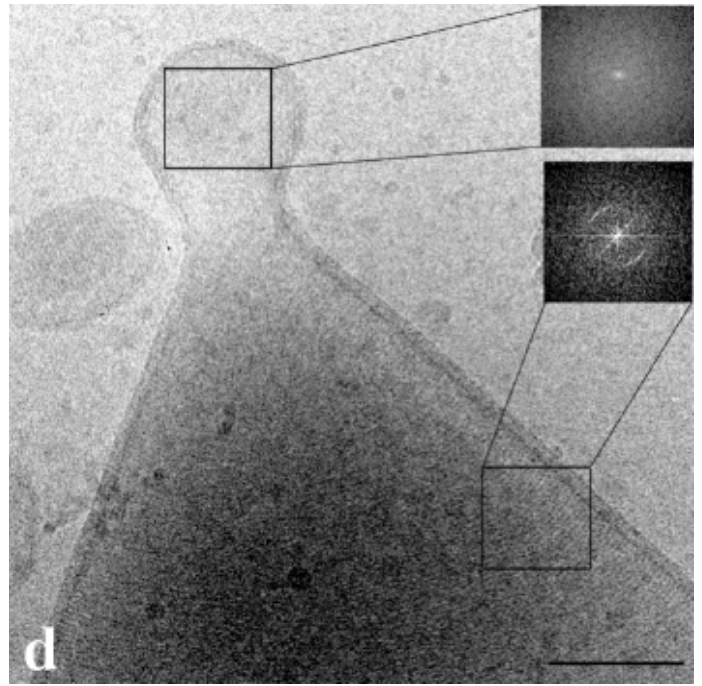

Figure 1

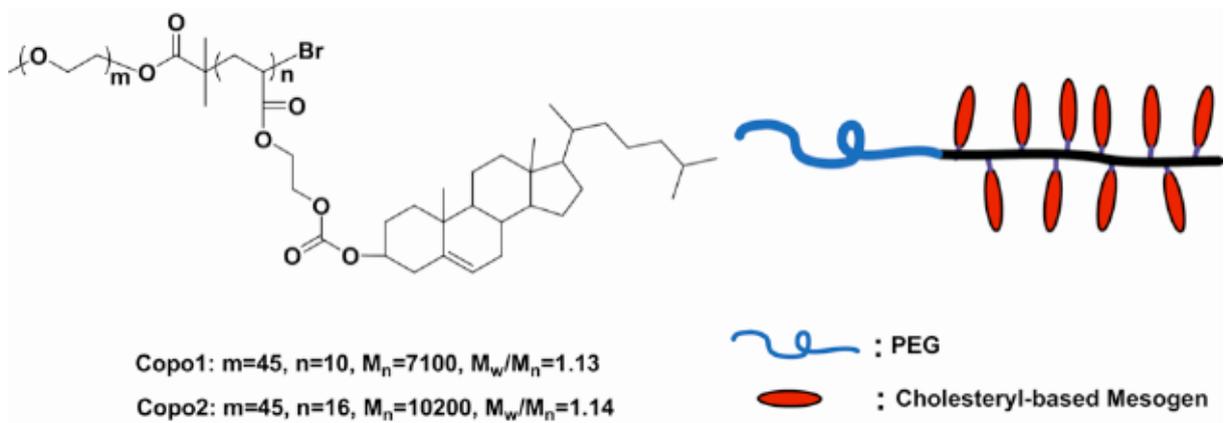

Figure 2

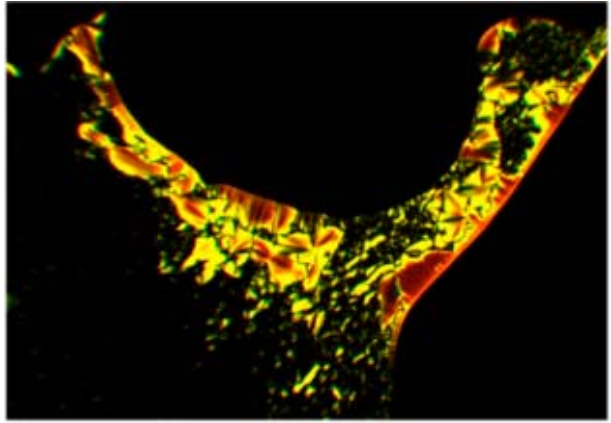

a

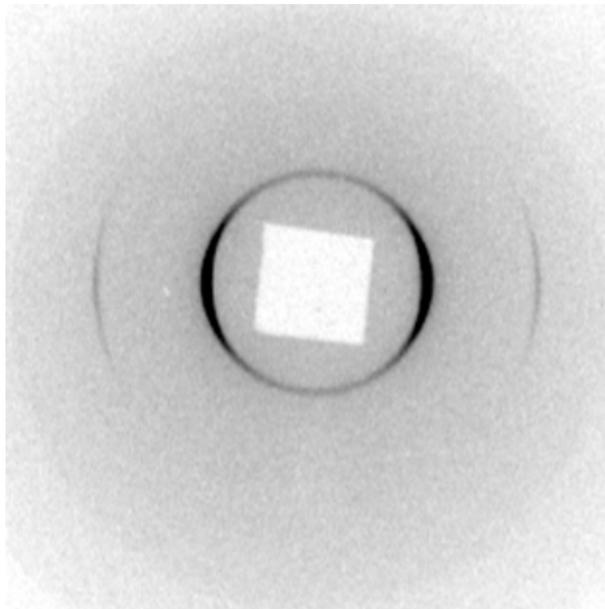

b

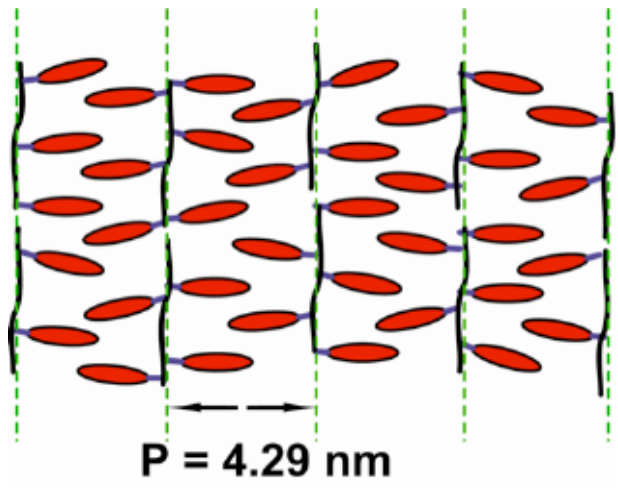

c

Figure 3

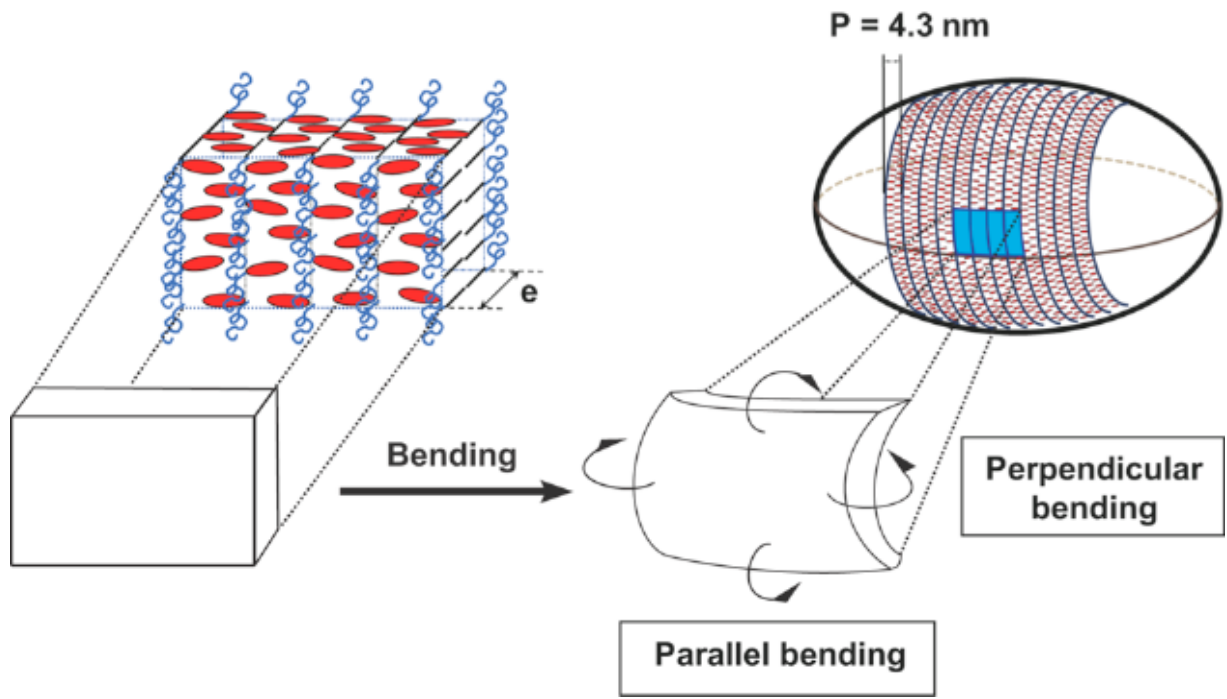

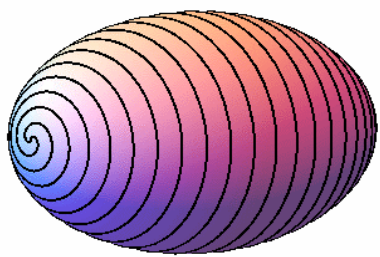

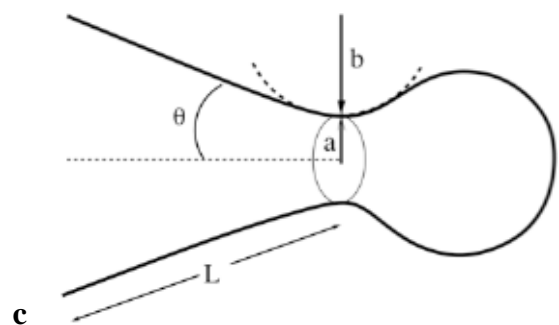

a

b c

Figure 4